\documentstyle[epsf]{l-aa}

\begin{document}

\thesaurus{04         
          (04.01.1;   
           04.03.1;   
           13.07.1)}  

\title{GRANAT/WATCH catalogue of cosmic gamma-ray bursts:
December 1989 to September 1994\thanks{Figure~\ref{fig2} is only available
in the on-line version of the paper; Table~1 is only available in
electronic form at the CDS via anonymous ftp to cdsarc.u-strasbg.fr
(130.79.128.5) or via http://cdsweb.u-strasbg.fr/Abstract.html}} 

\author{S.Y. Sazonov\inst{1,2}, 
R.A. Sunyaev\inst{1,2},
O.V. Terekhov\inst{1}, 
N. Lund\inst{3}, 
S. Brandt\inst{4} 
and A.J. Castro-Tirado\inst{5}
} 

\offprints{S.Y. Sazonov (sazonov@hea.iki.rssi.ru)}

\institute{Space Research Institute, Russian Academy of Sciences,
Profsoyuznaya 84/32, 117810 Moscow, Russia
\and 
Max-Planck-Institut f\"ur Astrophysik, Karl-Schwarzschildstr 1,
85740 Garching, Germany
\and 
Danish Space Research Institute, Juliane Maries Vej 30, DK 2100
Copenhagen \O. Denmark
\and 
Los Alamos National Laboratory, MS D436, Los Alamos, NM 87545, USA 
\and 
Laboratorio de astrof\'{\i}sica Espacial y F\'{\i}sica Fundamental
(LAEFF), INTA, P.O. Box 50727, 28080 Madrid, Spain
} 

\date{Received May 1997; accepted August 1997}

\maketitle

\markboth{S.Y. Sazonov et al.: GRANAT/WATCH catalogue of gamma-ray bursts}{}

\begin{abstract}
We present the catalogue of gamma-ray bursts (GRB) observed with the WATCH
all-sky monitor on board the GRANAT satellite during the period
December 1989 to September 1994. The cosmic origin of 95 bursts comprising the
catalogue is confirmed either by their localization with WATCH
or by their detection with other GRB experiments. For each burst
its time history and information on its intensity in the two energy
ranges 8--20 keV and 20--60 keV are presented. Most events show
hardening of the energy spectrum near the burst peak. In part of the
bursts an X-ray precursor or a tail is seen at 8--20~keV. We have
determined the celestial positions of the sources of 47 bursts. Their
localization regions (at $3\sigma$ confidence level) are equivalent in
area to circles with radii ranging from 0.2 to 1.6 deg. The burst
sources appear isotropically distributed on the sky on large angular scales. 

\end{abstract}

\section{Introduction}
From December 1989 to September 1994 the astrophysical observatory
GRANAT performed pointed observations of different celestial
regions. During that period, the X-ray instrument WATCH, a part of the
scientific payload of the observatory, was monitoring the whole of the
sky. WATCH is uniquely capable of precisely measuring the celestial
positions (the radius of the localization region is generally smaller
than 1~deg at the 3$\sigma$ confidence level) of short-lived hard
X-ray sources, which include GRBs. Another feature of the instrument
relevant to observations of GRBs is that its detectors are sensitive over an
X-ray energy range that reaches down to $\sim 8$~keV, the domain where the
properties of GRBs are known less than at higher energies.  

In this paper, we present the catalogue of GRBs detected with WATCH in
1989--1994. For nearly half of the events we 
have been able to determine the location of the burst source on the
celestial sphere. Earlier, a preliminary catalogue covering the WATCH
observations carried out before October 1992 was compiled by
\cite{castro-tirado94}. The new catalogue has been updated mainly
in the two aspects: 1) the bursts detected between October 1992 and September
1994 have been added, 2) more accurate positions have been
determined for many of the previously catalogued events due to the use
of more precise information on the spacecraft attitude and an improved
model of the instrument.   

\section{The WATCH experiment}
\begin{figure}[bt]
\epsfxsize=8.5cm \epsfbox[18 260 592 718]{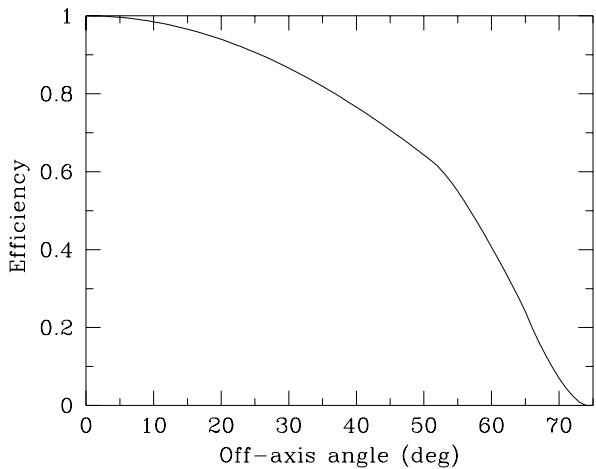}
\caption[]{Detector efficiency as a function of source off-axis angle (same
for both WATCH energy bands)}
\label{fig1}
\end{figure}

Four monitors WATCH (\cite{lund86}), designed at the Danish Space Research
Institute, are mounted on board the orbital observatory GRANAT. The field of
view of a monitor is a circle 74\degr\ radius. The scintillation
detector is made of alternating stripes of NaI and CsI. The full geometrical
area of the detector is 47~cm$^2$. The effective area of the detector is
dependent on the incident angle of the arriving photons as shown in
Fig.~\ref{fig1}: at small angles it declines as a cosine, starting at
$\sim 52$\degr\ more steeply, and for a source 65\degr\ off-axis it is one
fourth of that for an on-axis source. The fields of view of the four
instruments cover different quarters of the sky and partially overlap.
Because at the very beginning of the experiment one of the monitors went out
of order, $\sim 80$~\% of the sky can be simultaneously viewed under most
favourable conditions. The performance of the WATCH instrument is
based on the rotation modulation collimator principle. A point source,
provided it is bright enough for at least one rotation of the
modulation collimator ($\sim 1$~s), can be localized with an accuracy of
$\sim 50/n_{10}$~arcmin (at the $3 \sigma$ confidence level), where $n_{10}$
is the source signal in units of 10 standard deviations (\cite{brandt94}). The
detector count rate is recorded in two energy bands, the boundaries of which 
were reset several times during the mission and roughly correspond to 8--20~keV
and 20--60~keV.

\section{Registration of cosmic gamma-ray bursts}
WATCH possesses an on-board algorithm of triggering on burst events. A burst is
detected if an increase in the count rate of more than 6 standard
deviations above the background level has been registered at one of 12 sampling
times evenly covering on the logarithmic scale a range from 16~ms to
32~s. If an event is detected on a time scale shorter than 2~s,
its time history will be recorded with a resolution of 1~s, whereas
for slower bursts the standard integration time of either 7~s
or 14~s will be used. The purpose of the ground data analysis is then
to separate events of cosmic and different nature. These latter
include solar flares which differ from GRBs in their
generally softer spectra. It is also usually possible to establish
with WATCH that their incident direction is coincident with the
direction to the solar disc. In many cases burst events were generated
by accelerated charged particles. The time histories of such events
are different from those of GRBs. Although most non-cosmic
events can be reliably identified by their characteristic features,
the origin of a number of bursts remains unsettled. In this catalogue,
only those events are presented that either have been
localized with WATCH or are designated as cosmic in the catalogues of
other GRB experiments. The significance of the detection
of $\sim 25$~\% of the bursts included in the catalogue was not high
enough ($6 \sigma$) to generate a trigger on board. These events,
discovered already in the course of the ground analysis as a
significant ($> 4 \sigma$) increase in the count rate, are coincident
in time with GRBs observed by other experiments, mainly by
the BATSE instrument on the Compton Gamma-Ray Observatory (\cite{meegan96}).

\section{The catalogue}
A total of 95 GRBs have been included in the catalogue. The
information on these events is presented in Table~1. In the first
column of the table the names of the bursts are given, which were formed
using a common terminology: the first letter 'W' indicates the instrument's
name, WATCH, then follows the year, month and day of the detection of the
event. The burst may have a letter ('b' or 'c') appended to its name, if it is
the second or third burst detected in a day. The second column yields
the time (UT) of the trigger. In the subsequent columns the basic
characteristics of the bursts are given. As a measure of the burst 
duration we use the quantity $T_{90}$ which is the length of the
interval during which 90~\% (from 5~\% to 95~\%) of the total counts
from a burst was accumulated. The fluence and the peak energy flux of
the bursts were calculated in the two energy bands: 8--20~keV and
20--60~keV (spectral shape similar to that of the Crab Nebula was
assumed). When calculating these quantities we made a correction for
the aspect of the burst source, which is known for the majority of the
presented events from localizations with either WATCH or CGRO/BATSE
(\cite{meegan96}). However, for 18 bursts such information is not
available, hence we accepted for them as an estimate of the detector
geometrical efficiency (see Fig.~\ref{fig1}) its expectation value of 0.7. The
quoted errors for the fluxes are purely statistical ones, the
uncertainties due to unknown spectral shapes and source aspects have
not been considered.  The peak flux was calculated using the count
rate data with the best time resolution available for a given
event. Also, given in the table is the ratio of the fluences in the
20--60~keV and 8--20~keV energy bands, which characterizes the
hardness of the burst spectrum. Finally, the last column of the table
contains information on detections of the bursts by other experiments
that were in orbit during the period examined: KONUS/GRANAT,
PHEBUS/GRANAT, SIGMA/GRANAT, GINGA, BATSE/CGRO, COMPTEL/CGRO,
OSSE/CGRO, DMS, Mars Observer, PVO, ULYSSES, WATCH/EURECA and YOHKOH
(\cite{golenetskii91}; \cite{terekhov94}; \cite{terekhov95};
\cite{sunyaev93}; \cite{ogasaka91}; \cite{meegan96}; \cite{hanlon94};  
\cite{hurley94}; \cite{brandtetal94}).    

\subsection{Burst time histories}
In Fig.~\ref{fig2} the time histories of the detector count rate during the
bursts in the two WATCH energy bands are shown. For the longer bursts, the 
best time resolution available is either 7~s or 14~s. For the
shorter events, time histories of 1~s resolution are
presented. Finally, for the two shortest events W900404 and W930106,
which lasted less than 1~s, light curves obtained by integrating the
count rate over 20~ms intervals are presented. We 
note that these two time histories are necessarily distorted to some
degree by the modulation effect caused by the rotation of the
collimator. The background was in most cases estimated by averaging
the count rate over time intervals immediately before and after the
burst. When the background was strongly variable during the burst, we
used approximation of the count rate by polynomials of first or higher orders. 

\subsection{Burst durations}
Figure~\ref{fig3} shows the duration ($T_{90}$) distribution for the
bursts detected. It was obtained using those 89 out of 95 events
whose durations could be determined reliably. At least 5 of the 6
events excluded from the analysis are apparently short (shorter than a few
seconds), but they were too weak to force the on-board burst logic to trigger. 
Hence no count rate data with a good time resolution needed for
determining their durations is available. The mean burst duration is
66~s, and the maximum of the distribution lies in the interval 10 to 100~s.
Our sample contains 2 events shorter than 2~s and 7 events longer than
200~s. In other experiments, two classes of bursts were identified: of
duration shorter and longer than $\sim 2$~s (\cite{kouveliotou93}). It
was noticed that the energy spectra of the shorter bursts of type~1
were generally harder than those of the longer bursts of type~2
(\cite{kouveliotou93}; \cite{lestrade93}). The negligibly small number
of type~1 events in our catalogue compared to the previous results 
(\cite{meegan96}; \cite{terekhov95}) can be accounted for by at least two
selectional effects: 1) all bursts with detection significance less
than $6\sigma$ included in our duration sample are longer than 7~s, because
this value is the time resolution of the data, 2) WATCH is sensitive 
to X-ray photons with energies significantly below the effective energy of
the photons emitted in type~1 bursts.  

\setcounter{figure}{2}
\begin{figure}
\epsfxsize=8.5cm \epsfbox[18 260 592 718]{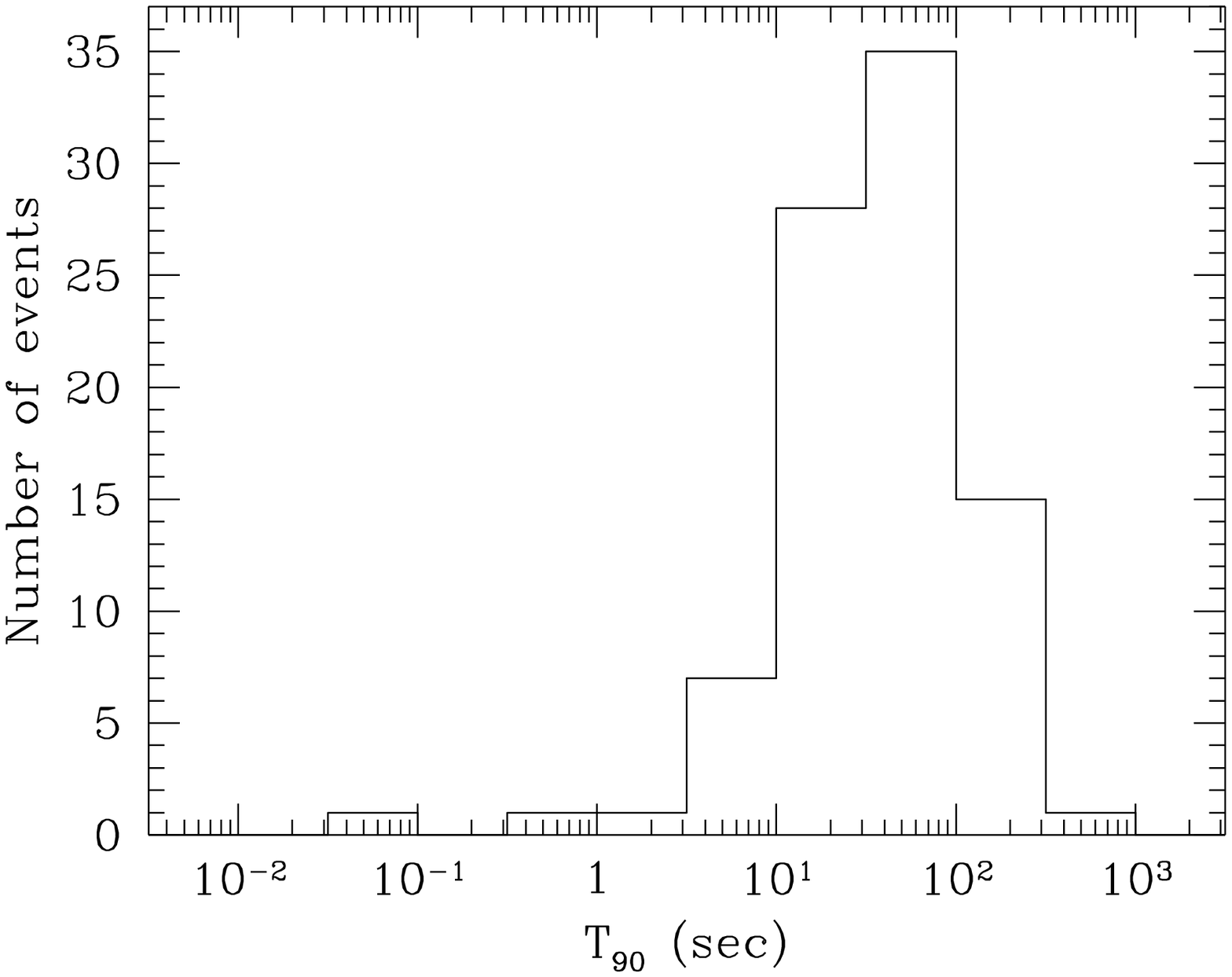}
\caption[]{Histogram of durations ($T_{90}$) for the WATCH bursts}
\label{fig3}
\end{figure}

\subsection {Burst spectral hardness and evolution}
\begin{figure}
\epsfysize=8.5cm \epsfbox{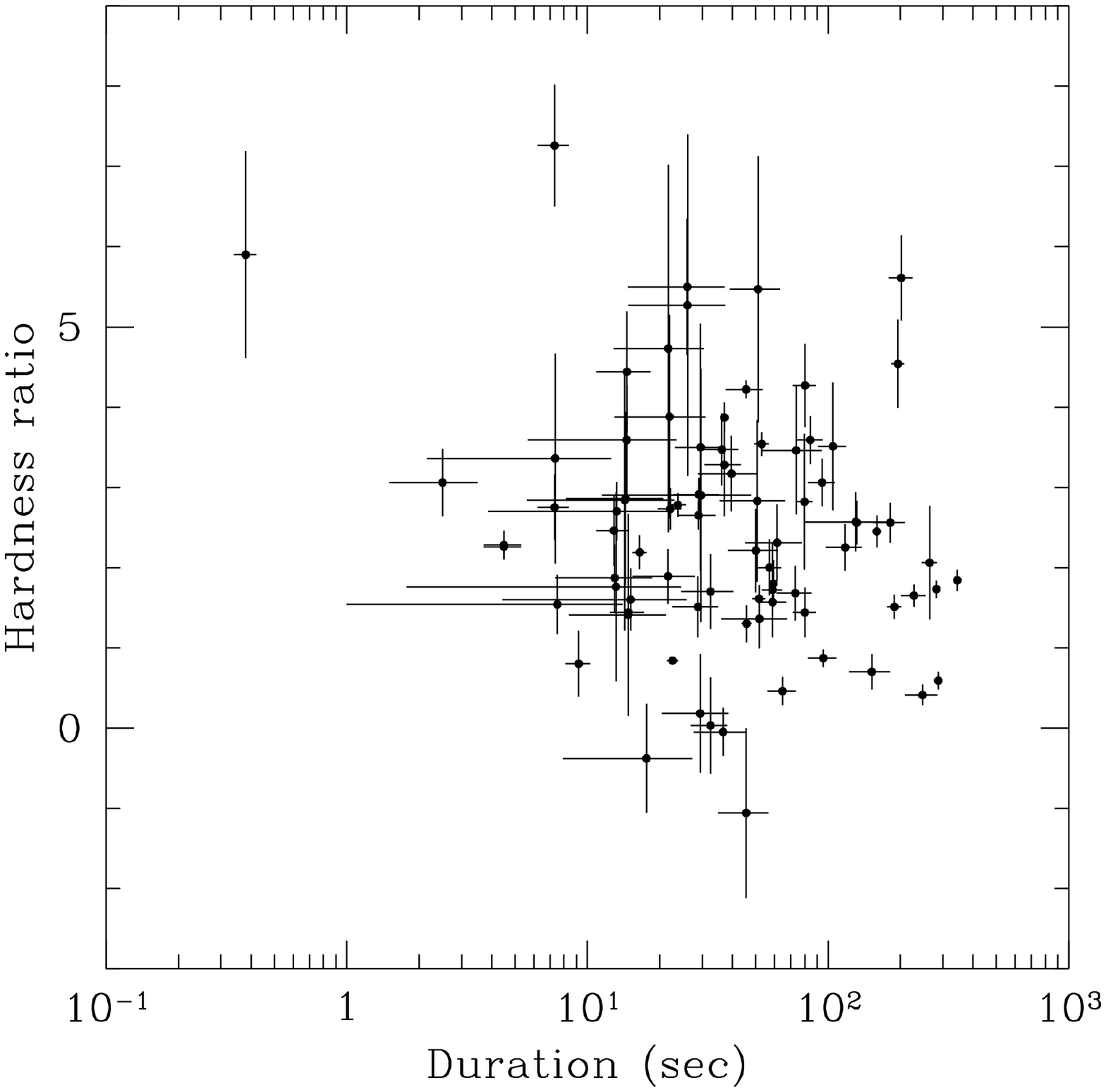}
\caption[]{Burst hardness ratios (fluence over the 20--60~keV band divided by
that over the 8--20~keV band) vs. burst durations}
\label{fig4}
\end{figure}

The spectral information on the emission produced during GRBs provided by
WATCH is limited to the count rates in the two 
energy ranges 8--20~keV and 20--60~keV. Hence to describe the spectra
of the observed bursts we calculated the ratios of the burst energy
fluxes in the higher and lower bands. As follows from the
cross-correlation diagram presented in Fig.~\ref{fig4}, there is no clear 
dependence of burst overall hardness ratio (the 20--60~keV fluence
divided by the 8-20~keV fluence) on burst duration. This graph gives further
support to the above statement that most of the bursts in the WATCH
catalogue are of the same type.

\begin{figure}
\epsfysize=8.5cm \epsfbox{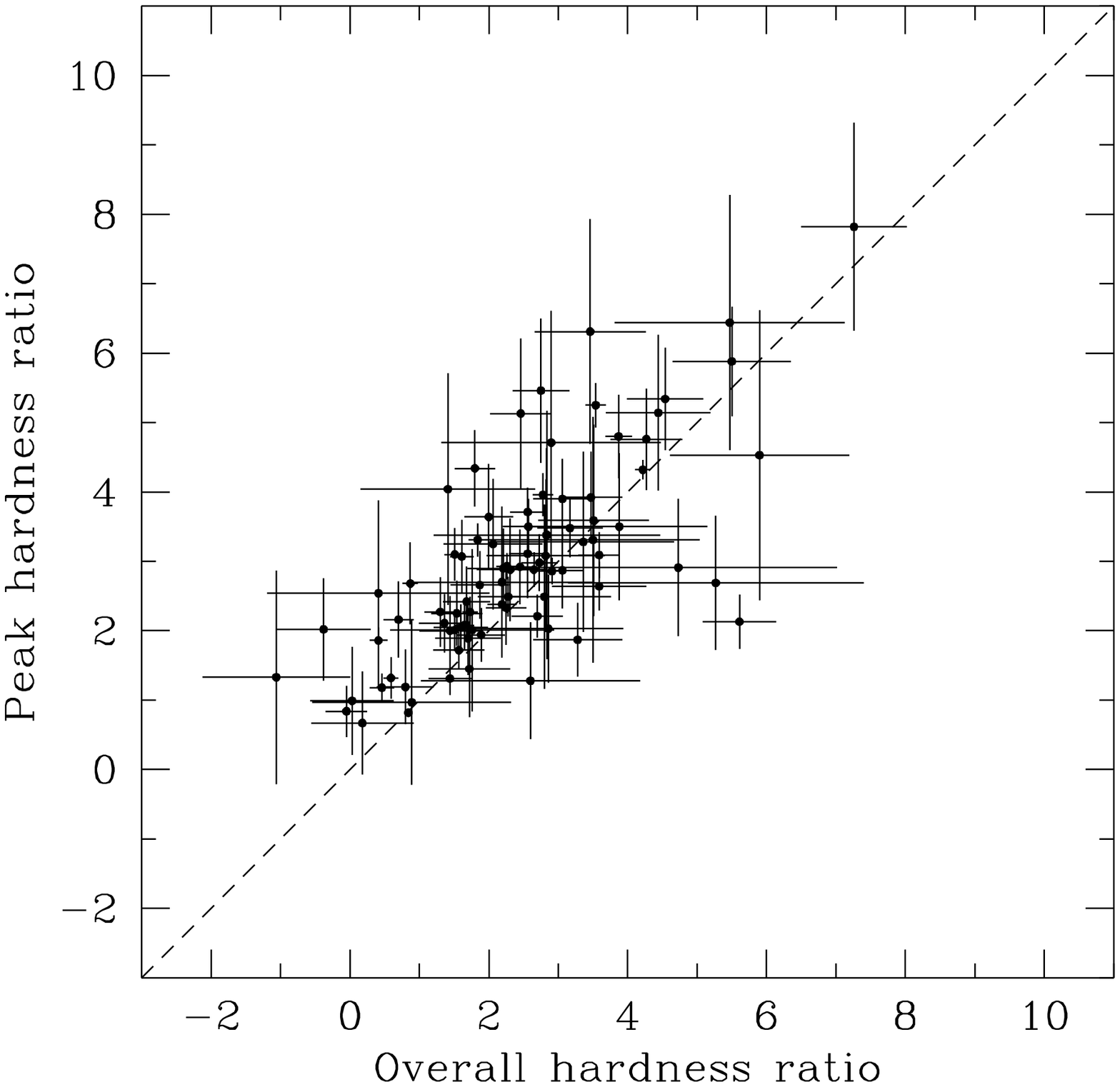}
\caption[]{Burst peak hardness ratios vs. burst overall hardness ratios.
The dashed line indicates the case of equality of these two quantities}
\label{fig5}
\end{figure}

The WATCH observations illustrate that the energy spectra are usually
not constant but evolve throughout bursts, the typical situation being
that the spectra during the burst rise and decay phases are softer
than that at the peak of the event. This is demonstrated by Fig.~\ref{fig5},
where burst peak hardness ratio, i.e. the ratio of the peak energy fluxes in
20--60~keV and 8--20~keV (in some bursts the two flux maxima are not
coincident in time), is shown as a function of burst overall hardness ratio.
It can be seen that the former is larger than the latter for the majority of
events, reflecting the fact that bursts generally have ``sharper''
profiles in the harder energy band. In 13 bursts (Table~\ref{tab2}) this
spectral evolution reveals itself especially distinctly as a
significant activity observed only at 8--20~keV either preceding or 
following the hard X-ray event (see the corresponding time histories
in Fig.~\ref{fig2}, for discussion on part of these events with occurences
before October 1992 see also \cite{castro-tirado94} and Castro-Tirado
et al. (1994)). Similar X-ray precursor and tail activities have been
observed before in a number of bursts by a few space-flown GRB instruments
sensitive to medium or soft X~rays (\cite{murakami91}). In the observations
carried out with the GRB detector on board the GINGA satellite,
which had a low-energy cut-off at as low as 1.5~keV, such X-ray-active events
accounted to about one third of the total bursts detected
(\cite{murakami92}). The photon spectra measured with GINGA at 1 to
10~keV during both the burst X-ray precursor and tail could be
approximated by a black-body model with temperatures between 1 and
2~keV, indicating that the emission mechanism at these burst phases
may be thermal at variance with the apparently non-thermal emission during the
main gamma-ray event.

\setcounter{table}{1}
\begin{table}
\caption[]{Gamma-ray bursts with precursor or tail X-ray activity}
\begin{tabular}{ll}
\noalign{\smallskip}
\hline
\noalign{\smallskip}
Burst name & Type of X-ray activity\\
\hline
\noalign{\smallskip}
W900222  & precursor and tail\\
W900708  & precursor and tail\\
W900901  & tail\\
W910817  & tail\\
W911209  & precursor\\
W920718  & precursor\\
W920723b & tail\\
W920903  & tail\\
W920903b & precursor\\
W920925  & precursor\\
W921013b & tail\\
W921022  & tail\\
W930705  & tail\\
\noalign{\smallskip}
\hline
\end{tabular}
\label{tab2}
\end{table}

\subsection{Distribution of bursters in space}
For describing the distribution of GRB sources in space, the
$V/V_{max}$ statistics (\cite{schmidt88}) is widely used. In this
statistics, for each burst detected according to the set criterium one
calculates the quantity $V/V_{max}$ by the formula:
$V/V_{max}=(C_{max}/C_{min})^{-3/2}$, where $C_{min}$ is the minimum 
count rate to satisfy the burst detection criterion, and $C_{max}$ is
the maximum count rate during the burst. If the bursters are
homogeneously distributed in space, $V/V_{max}$ will be uniformly
distributed in the interval (0,1), and its expectation will be 0.5. 

We have carried out a $V/V_{max}$ test on the WATCH GRB catalogue.
Calculating the values of the parameters $C_{min}$ and $C_{max}$, 
in order to have a homogeneous sample, we used the count rate data
with 14~s integration time, for data of such resolution exist for
all of the events. Of crucial importance in our case is the definition
of $C_{min}$, since the inclusion of many bursts into the catalogue was
dependent on observations of the event by other experiments. We have
chosen the criterion that the peak count rate must exceed the
background by at least $\sim 15\sigma$ (this limit very roughly
corresponds to a flux threshold of $\sim
3\times10^{-7}$~erg~cm$^{-2}$~s for the WATCH sensitivity energy range
and a typical trigger time scale of 8~s), for it proves that virtually
all cosmic bursts meeting it can be registered with WATCH
independently, through localization. The average value of $V/V_{max}$
for 43 so selected events is $0.54\pm0.04$. This implies that the
burst sources are homogeneously distributed within the WATCH sampling
distance. The WATCH instrument is not sensitive enough to make it
possible for us to analyze the distribution of more distant bursters,
a deficit of which has been observed in a number of burst experiments
(\cite{meegan96}; \cite{terekhov95}).  

\subsection{Burster locations in the sky} 
The location of the source of a burst can be found from the WATCH data if
the following three requirements are met: 1) the burst is strong
enough, 2) it lasts longer than $\sim$~1 rotation of the modulation
collimator, and 3) the modulation pattern used for localization 
is not significantly distorted by the presence in the light curve of bright
details on time scales shorter than the rotation period of the 
collimator. The presence of other bright sources in the field of view can
possibly make the procedure of deriving the source position unreliable even
when the above conditions are satisfied. Therefore we 
considered a source localized only if the same position was resulted from two
statistically independent modulation patterns. These patterns may belong to
different time intervals, different energy ranges, or, in rare cases,
different phase intervals of the same modulation pattern. For several weak
bursts an additional independent verification was obtained through comparing
the positions provided by WATCH and BATSE.  

We have succeeded in localizing the sources of 47 bursts (Table~\ref{tab3}).
The statistical uncertainty of position determination is inversely proportional
to the significance of source detection and varies between 7~arcmin and
1.5~deg for the localized bursts (the radius of a circle with an area equal
to the area of the $3 \sigma$ confidence region), the localization region
being an ellipse somewhat contracted along the source off-axis angle
$\theta$. In preparation of this catalogue special efforts were made to
decrease the influence of various systematic effects on burst
localizations. Significant progress in this direction has now been   
achieved, in the first instance due to the use of information provided
by the star tracker of the SIGMA telescope for determination of the
attitude of the spacecraft at the times of bursts. Besides, on the
basis of an ample archive of WATCH data on localizations of bright
persistent X-ray sources, we have found more accurate values for some
of the parameters relevant to the instrument, including the mounting
angles defining the orientation of the WATCH detectors with respect to
the SIGMA star tracker. Unfortunately, the star tracker was at times
off during WATCH observations, and for 16 bursts we thus were bound to
use other information resources to calculate the attitude, namely
readings of the navigational instruments of the spacecraft 
and the knowledge of the celestial positions of bright X-ray 
sources that are always present in the field of view and can be localized
with the WATCH detectors. This leaves an irremovable uncertainty $\sim
0.5$\degr\ resulted from the rapid ($\sim~30$~min period) and virtually
unpredictable wobbling of the spacecraft attitude. For the 31 positions that
were calculated using precise navigational information we
conservatively estimate the remaining systematic error at
0.2\degr. This uncertainty is mainly due to the not complete accounting
for various physical phenomena in the instrument mathematical model
currently used, in particular the dependence of the instrument's
positional response on the energy spectrum of the incident
radiation. The cumulative localization uncertainties given in the last
column of Table~\ref{tab3} were calculated by summing (in quadrature) the
statistical ($3 \sigma$) and estimate systematic errors. Although we
have already arrived at the state that for most of the localized burst
sources it is the statistical error that mostly contributes to the location
uncertainty, a cooperative effort between IKI and DSRI is now in progress to
further improve the modelling of the instrument, which we expect will
eventually enable further reducing of the localization regions of several
stronger bursts.      

\begin{figure*}
\epsfxsize=14.5cm \epsfbox{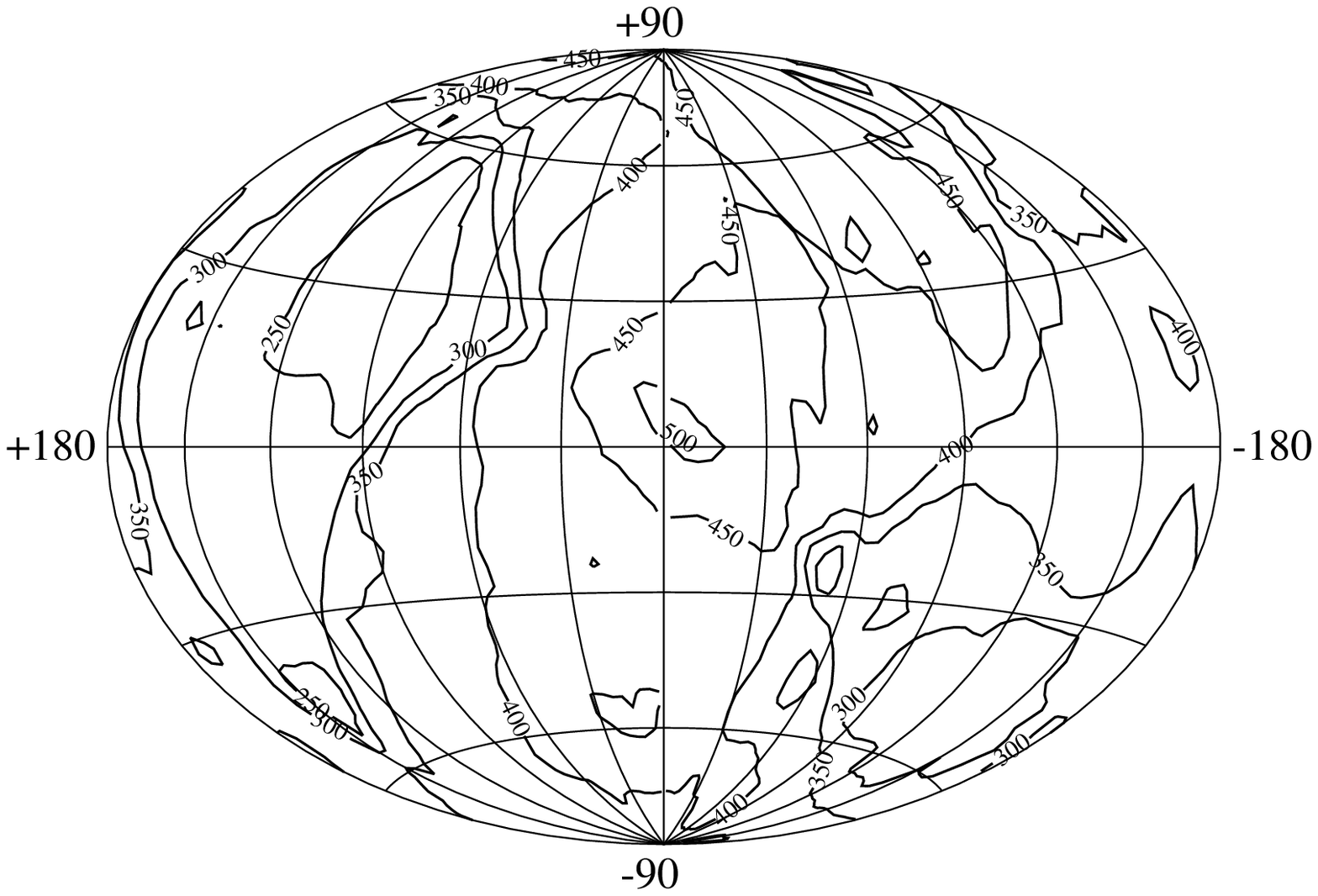}
\caption[]{Sky exposure map in Galactic coordinates constructed
from the data of WATCH observations in 1989--1994. Exposure time is
measured in days}
\label{fig6}
\end{figure*}

When analyzing a celestial distribution of sources, it is necessary to know
how long different regions of the sky have been monitored. We have compiled
from the WATCH data an exposure map, which is shown in Galactic
coordinates in Fig.~\ref{fig6}. Calculating this map, in cases when a sky
region had been observed simultaneously by two or 
three WATCH detectors, the corresponding time interval was appended
only once. We considered an area of the sky being in sight of the
instrument if it was not more than 65\degr\ off axis. The maximum of
the exposure map of 510~days is located in the vicinity of the
Galactic center ($l= -7$\degr, $b= 3$\degr), and its minimum of
218~days has coordinates $l=87$\degr, $b=29$\degr. The exposure time
averages 372~days over the celestial sphere, which corresponds to an
all-sky monitoring efficiency of 21~\%. 

\begin{figure*}
\epsfxsize=14.5cm \epsfbox{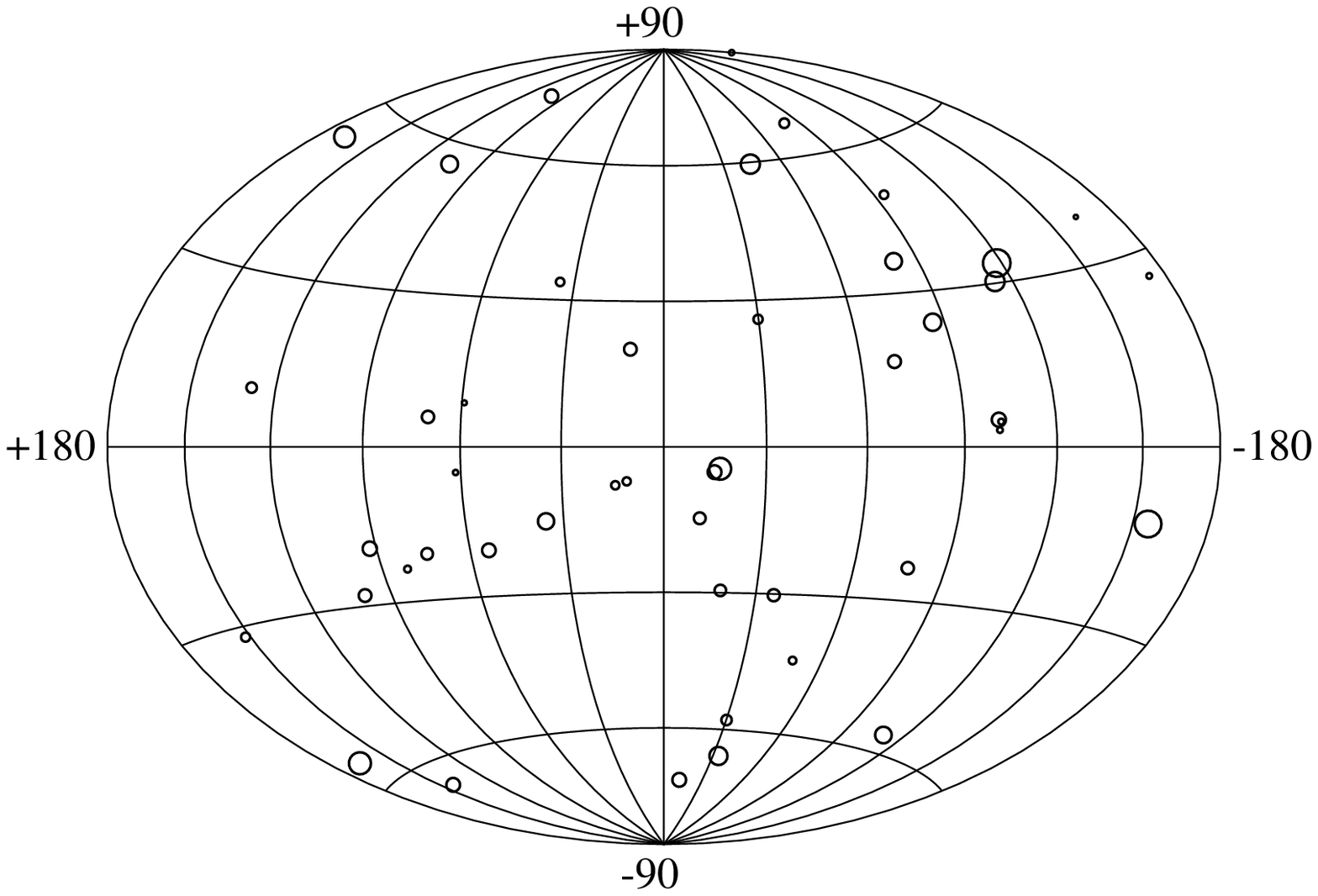}
\caption[]{Positions of 47 burst sources in the sky in Galactic
coordinates. The circles shown are the twice-zoomed real localization
circles}
\label{fig7}
\end{figure*}

In Fig.~\ref{fig7}, a celestial map of the positions of the burst sources in
Galactic coordinates is presented. In order to check the angular distribution
of the bursts for possible large-scale anisotropies, we have calculated its 
dipole and quadrupole moments relative to the Galactic center and the
Galactic plane, respectively. Upon correction for exposure time
(terms are summed with weights inversely proportional to the position
exposure) the dipole moment $<\cos\theta>=0.10\pm0.08$ ($\theta$ is
the source angular distance from the Galactic center), the quadrupole
moment $<\sin^{2}b-1/3>=-0.01\pm 0.04$ ($b$ is the source galactic
latitude). Therefore our observations are consistent with an isotropic
distribution of the burst sources on the sky, in agreement with the
corresponding BATSE result (\cite{meegan96}). We
carried out similar calculations for samples of bursts selected out by
various attributes. We have not found any significant deviations from
isotropy, in particular, the distribution of the sources of the
stronger bursts is apparently isotropic. Note that the dipole and
quadrupole moments calculated using the 32 events of the first WATCH
catalogue of GRBs are: $<\cos\theta>= 0.22\pm0.10$,
$<\sin^{2}b-1/3>=-0.04\pm 0.05$, respectively (\cite{castro-tiradoetal94}).
Thus, the tendency for bursters to concentrate towards the Galactic center,
evident at a significance of $2\sigma$ in the first catalogue, is not
confirmed by the new data obtained since October 1992.

\begin{table*}
\caption[]{GRANAT/WATCH localizations of gamma-ray bursts}
\begin{tabular}{lcccccc}
\noalign{\smallskip}
\hline
\noalign{\smallskip}
Burst & $\alpha$ (2000.0) & $\delta$ (2000.0) & $l$ & $b$ &  
$3 \sigma$ stat. error & Total error \\   
name & (\degr) & (\degr) & (\degr) & (\degr) & (\degr) & (\degr)  \\
\hline
\noalign{\smallskip}
W900118  &  174.68 & -44.32 & 289.46 & ~16.64 & 0.54 & 0.73 \\   
W900123b &  357.19 & -38.56 & 347.89 & -72.61 & 0.60 & 0.78 \\
W900126  &  131.15 & -37.79 & 258.76 & ~~3.09 & 0.24 & 0.32 \\
W900222  &  336.73 & ~34.83 & ~92.09 & -19.25 & 0.79 & 0.81 \\
W900708  &  185.91 & ~30.62 & 181.40 & ~82.99 & 0.22 & 0.30 \\
W900708b &  252.79 & ~16.20 & ~34.92 & ~33.74 & 0.43 & 0.48 \\
W900901  &  276.31 & -45.18 & 349.28 & -14.59 & 0.45 & 0.67 \\
W900925  &  133.14 & -36.72 & 258.92 & ~~5.00 & 0.77 & 0.80 \\
W900929  &  169.68 & ~-6.48 & 265.83 & ~49.59 & 0.44 & 0.49 \\
W901009  &  348.61 & ~30.40 & ~99.28 & -27.98 & 0.51 & 0.72 \\
W901116  &  ~39.93 & ~24.97 & 151.95 & -31.73 & 0.48 & 0.52 \\
W901121  &  ~30.39 & ~72.40 & 128.24 & ~10.26 & 0.55 & 0.59 \\
W901219  &  348.62 & -54.00 & 329.84 & -57.77 & 0.55 & 0.59 \\
W910122  &  297.48 & -71.23 & 324.06 & -30.26 & 0.66 & 0.69 \\
W910219  &  212.94 & ~58.54 & 104.43 & ~55.61 & 0.93 & 0.95 \\
W910310  &  184.10 & ~~6.38 & 279.48 & ~67.64 & 0.24 & 0.55 \\
W910627  &  199.60 & ~-2.60 & 316.32 & ~59.57 & 1.08 & 1.09 \\
W910821  &  353.08 & -72.01 & 311.26 & -43.81 & 0.38 & 0.43 \\
W910927  &  ~49.70 & -42.72 & 250.24 & -56.37 & 0.80 & 0.94 \\
W911016  &  297.37 & ~-4.71 & ~35.35 & -15.05 & 0.78 & 0.92 \\
W911202  &  171.97 & -22.59 & 278.81 & ~36.34 & 0.80 & 0.94 \\
W911209  &  261.92 & -44.19 & 345.18 & ~-5.15 & 0.60 & 0.78 \\
W920210  &  154.15 & ~47.89 & 167.78 & ~53.51 & 1.08 & 1.19 \\
W920311  &  132.25 & -36.39 & 258.20 & ~~4.65 & 0.26 & 0.33 \\
W920404  &  323.07 & ~22.53 & ~73.92 & -20.85 & 0.63 & 0.66 \\
W920714  &  221.43 & -30.75 & 330.15 & ~26.01 & 0.48 & 0.52 \\
W920718  &  ~21.37 & ~-3.36 & 143.31 & -64.88 & 0.75 & 0.78 \\
W920718b &  296.17 & -55.95 & 341.69 & -29.51 & 0.62 & 0.65 \\
W920720  &  145.67 & -11.20 & 246.46 & ~30.31 & 1.07 & 1.09 \\
W920723b &  287.08 & ~27.33 & ~59.26 & ~~8.69 & 0.19 & 0.28 \\
W920814  &  259.83 & -45.17 & 343.53 & ~-4.47 & 1.13 & 1.24 \\
W920902  &  279.08 & -22.81 & ~10.87 & ~-7.04 & 0.41 & 0.46 \\
W920903  &  295.87 & ~35.46 & ~70.02 & ~~5.81 & 0.67 & 0.70 \\
W920903b &  301.54 & ~22.59 & ~61.53 & ~-5.05 & 0.22 & 0.30 \\
W920925  &  201.11 & ~42.20 & 100.96 & ~73.49 & 0.73 & 0.76 \\
W920925c &  330.80 & ~25.48 & ~81.64 & -23.60 & 0.34 & 0.39 \\
W921013  &  ~87.97 & ~~1.93 & 204.27 & -12.31 & 1.50 & 1.51 \\
W921013b &  117.71 & ~33.41 & 186.99 & ~26.20 & 0.25 & 0.32 \\
W921022  &  254.43 & ~-9.64 & ~10.21 & ~19.97 & 0.52 & 0.72 \\
W921029  &  ~35.82 & ~-0.42 & 166.28 & -55.37 & 1.23 & 1.25 \\
W930612  &  109.24 & -71.20 & 282.48 & -23.56 & 0.68 & 0.70 \\
W930703  &  311.06 & ~~8.04 & ~54.10 & -20.65 & 0.59 & 0.77 \\
W930706  &  281.42 & -20.18 & ~14.23 & ~-7.83 & 0.42 & 0.47 \\
W940419  &  358.82 & -48.19 & 326.63 & -66.27 & 1.00 & 1.02 \\
W940701  &  145.67 & ~-6.15 & 241.88 & ~33.54 & 1.54 & 1.56 \\
W940703  &  133.20 & ~28.11 & 197.08 & ~37.69 & 0.12 & 0.24 \\
W940907  &  161.42 & -31.81 & 273.93 & ~23.88 & 0.84 & 0.98 \\
\noalign{\smallskip}
\hline
\end{tabular}
\label{tab3}
\end{table*}

\section{Summary}
The WATCH GRB catalogue provides a set of 47 burster
positions, 39 of which have a total (statistical plus systematic)
uncertainty of less than 1~deg. For 13 events error boxes with
radii smaller than 30~arcmin are now available. The WATCH sample thus
contributes to the currently available list of moderately accurate
GRB positions a number of locations comparable to that accumulated by the
interplanetary satellite networks and never before obtained with any
stand-alone instrument. We therefore hope that the new data presented
here will be useful for burster counterpart searches in different
energy ranges as well as for studying possible correlations in GRB positions.

The WATCH GRBs appear distributed both isotropically on
the celestial sphere and homogeneously in space. These two results
seem to be consistent with the implications from the third BATSE
catalogue, as WATCH is about an order of magnitude less sensitive than
the large-area detectors of BATSE, and the brighter bursts in the
BATSE catalogue also show both isotropy and homogeneity (\cite{meegan96}). 

The light curves of most bursts observed by WATCH show hardening of
the energy spectrum near the burst maxima. Several bursts demonstrate a
significant 8--20~keV activity in the absence of hard X-ray flux either
before or after the GRB. To all appearances, these X-ray events accompanying
gamma-ray bursts are higher energy manifestations of the soft X-ray
precursors and tails observed at 1.5--10~keV by GINGA.

\begin{acknowledgements}
We thank all specialists involved in the GRANAT/WATCH experiment,
in particular: the staffs of the Lavochkin Association, the Evpatoria
Deep Space Control Center and the group of B.S. Novikov maintaining
the instrument operations; the group of A.V. Dyachkov at IKI RAN
processing telemetry data; N.G. Havenson and E.M. Churazov for their
assistance in determining the spacecraft attitude. The participation
of the russian co-authors in this project was supported by Russian
Basic Research Foundation grant No. 95-02-05938 and INTAS grant
93-3364. SS acknowledges the hospitality of the Max-Planck-Institut
f\"ur Astrophysik where the final manuscript of this paper was finished.   
\end{acknowledgements}


\setcounter{table}{0}
\begin{table*}
\caption[]{Basic information on GRANAT/WATCH gamma-ray bursts}
\begin{tabular}{llr@{$\pm$}lr@{$\pm$}lr@{$\pm$}lr@{$\pm$}lr@{$\pm$}lr@{$\pm$}lll}
\noalign{\smallskip}
\hline
\noalign{\smallskip}
Burst & Start & \multicolumn{2}{c}{$T_{90}$} & \multicolumn{4}{c}
{\mbox{Fluence (10$^{-7}$ erg/cm$^{2}$})} & \multicolumn{2}{c} {Hardness} &
\multicolumn{4}{c} {\mbox{Peak flux (10$^{-7}$ erg/cm$^{2}$/s})} & Rem & Confirm$^{\rm a}$\\
\cline{5-8} \cline{11-14}  
name & (UT) & \multicolumn{2}{c} {(s)} & \multicolumn{2}{c} {8--20 keV} &
\multicolumn{2}{c} {20--60 keV} & \multicolumn{2}{c} {ratio} &
\multicolumn{2}{c} {8--20 keV} & \multicolumn{2}{c} {20--60 keV} & & \\    
\hline
\noalign{\smallskip}
W900118  &  17 40 06 &  227    &   27    &  102.1  &    4.9  &  169    &    11   &    1.65 &    0.14 &    0.87 &    0.10 &    1.81 &    0.23 &   & \\
W900120  &  20 37 51 &   61    &   16    &   28.3  &    4.2  &   65.3  &    9.2  &    2.31 &    0.48 &    0.67 &    0.13 &    1.91 &    0.30 & b & K\\
W900123  &  01 42 17 &    7.5  &    6.5  &    8.6  &    1.1  &   13.2  &    2.6  &    1.54 &    0.37 &    0.63 &    0.09 &    1.41 &    0.21 & b & K\\
W900123b &  18 44 31 &   15    &   10    &   25.3  &    3.5  &   40.6  &    8.1  &    1.60 &    0.39 &    1.01 &    0.10 &    2.08 &    0.26 &   & \\
W900126  &  18 04 22 &   22.6  &    1.2  &   79.7  &    1.5  &   67.3  &    2.9  &    0.84 &    0.04 &   18.95 &    0.52 &   15.60 &    0.79 &   & G\\
W900222  &  11 56 34 &  246    &   38    &  110.8  &    5.2  &   44    &    14  &    0.41 &    0.13 &    1.73 &    0.10 &    3.21 &    0.27  &   & Ph\\
W900308  &  09 39 09 &   29    &   18    &    4.7  &    2.2  &   13.7  &    3.8  &    2.9  &    1.5  &    0.16 &    0.06 &    0.76 &    0.11 & b & KPh\\
W900404  &  17 54 35 &    0.38 &    0.04 &    1.04 &    0.20 &    6.14 &    0.56 &    5.9  &    1.2  &    7.5  &    2.9  &   34.2  &    8.2  & b & PhS\\ 
W900413  &  10 11 30 &   37    &   15    &   \multicolumn{2}{c}{--}  &
\multicolumn{2}{c}{--} &  \multicolumn{2}{c}{--} &  \multicolumn{2}{c}{--}
&  \multicolumn{2}{c}{--}  & b & Ph\\
W900618  &  23 12 24 &   73    &   20    &   17.2  &    3.3  &   59.5  &    7.2  &    3.46 &    0.80 &    0.46 &    0.11 &    2.93 &    0.26 & b & Ph\\
W900708  &  11 28 10 &  181    &   27    &  108.9  &    8.1  &  279    &   17    &    2.56 &    0.25 &    2.57 &    0.20 &    9.49 &    0.48 &   & \\
W900708b &  18 44 54 &   94    &   12    &   59.9  &    5.0  &  183.2  &    8.8  &    3.06 &    0.30 &    0.87 &    0.14 &    2.48 &    0.26 &   & \\
W900901  &  01 33 40 &  195    &   12    &   38.9  &    4.3  &  176.9  &    8.6  &    4.54 &    0.55 &    0.74 &    0.09 &    3.99 &    0.22 &   & \\
W900925  &  20 25 22 &   45.9  &    2.2  &   33.6  &    3.0  &   43.6  &    6.5  &    1.30 &    0.23 &    3.22 &    0.48 &    7.3  &    1.1  &   & Ph\\
W900929  &  01 49 26 &  159.4  &    2.6  &   76.8  &    4.6  &  188    &   10    &    2.45 &    0.20 &    3.22 &    0.45 &    9.3  &    1.1  &   & Ph\\
W901009  &  00 15 48 &   14.6  &    3.7  &   11.7  &    1.8  &   52.0  &    3.7  &    4.44 &    0.75 &    1.85 &    0.37 &    9.55 &    0.93 &   & \\
W901015  &  08 44 46 &   \multicolumn{2}{c}{--} &   12.8  &    2.9  &   28    &   15    &    2.1  &    1.3  &    0.67 &    0.11 &    1.80 &    0.64 & b &  Ph\\
W901027  &  04 21 01 &   13.2  &    9.3  &   17.7  &    1.8  &   47.8  &    4.3  &    2.70 &    0.37 &    1.04 &    0.10 &    2.30 &    0.24 & b & PhS\\
W901112  &  14 57 41 &   21.9  &    8.9  &    4.7  &    1.3  &   18.5  &    2.9  &    3.8  &    1.2  &    0.37 &    0.10 &    1.31 &    0.21 & b & PhU\\
W901116  &  01 42 12 &   14.5  &    8.9  &   11.3  &    1.8  &   40.8  &    3.7  &    3.59 &    0.68 &    1.21 &    0.12 &    3.22 &    0.26 &   & Ph\\
W901121  &  18 07 11 &  188    &   12    &   61.5  &    3.7  &   92.8  &    7.0  &    1.51 &    0.15 &    0.65 &    0.07 &    2.00 &    0.13 &   & PhU\\
W901219  &  21 52 43 &   39    &   10    &   46.1  &    5.6  &  146    &   12    &    3.17 &    0.47 &    2.30 &    0.22 &    8.00 &    0.54 & c & \\
W910117  &  00 58 13 &   51    &   12    &   18.6  &    3.4  &  102    &   24    &    5.4  &    1.6  &    0.73 &    0.14 &    4.6  &    1.0  & b & PhU\\ 
W910122  &  15 14 00 &   94.8  &    7.1  &   \multicolumn{2}{c}{--}  &    \multicolumn{2}{c}{--} &  \multicolumn{2}{c}{--} &  \multicolumn{2}{c}{--}  &  \multicolumn{2}{c}{--}  &   & GPPhSU\\
W910219  &  11 45 24 &   59.2  &    1.9  &   69.3  &    2.9  &  124    &   19    &    1.80 &    0.29 &    7.22 &    0.51 &   31.2  &    3.3  &   & PU\\
W910310  &  13 02 03 &   37.10 &    0.80 &   77.8  &    3.1  &  301.4  &    8.3  &    3.87 &    0.19 &    4.71 &    0.50 &   22.6  &    1.4  &   & PU\\ 
W910425  &  00 38 05 &  104    &   13    &   57    &   10    &  203    &   30    &    3.51 &    0.80 &    1.15 &    0.34 &    4.13 &    0.99 &   & BCPU\\
W910517  &  05 02 38 &   37.1  &    6.4  &   59.0  &    4.3  &  193    &   34    &    3.28 &    0.64 &    3.06 &    0.20 &    5.7  &    1.5  & b,c & PPhSU\\
W910627  &  04 29 23 &   14.8  &    2.4  &   48.1  &    3.1  &   69    &   20    &    1.44 &    0.44 &   10.38 &    0.89 &   20.8  &    4.9  &   & BCPPhU\\
W910717  &  04 33 06 &    9.2  &    1.1  &   27.8  &    2.5  &   22    &   11    &    0.80 &    0.41 &    7.09 &    0.94 &    8.4  &    3.6  &   & BGPPhU\\
W910717b &  13 07 23 &   14.7  &    6.4  &   15.9  &    3.9  &   22    &   19    &    1.4  &    1.2  &    1.04 &    0.27 &    4.2  &    1.3  & b & U\\
W910721  &  19 30 17 &   29.4  &    9.0  &   28.4  &    4.7  &    5    &   21    &    0.18 &    0.74 &    1.76 &    0.30 &    1.1  &    1.2  &   & BU\\ 
W910814  &  19 14 38 &   \multicolumn{2}{c}{--}  &  \multicolumn{2}{c}{--}
&    \multicolumn{2}{c}{--} &  \multicolumn{2}{c}{--} &  \multicolumn{2}{c}{--}  &  \multicolumn{2}{c}{--} & d & BCGOPPhSU\\
W910815  &  12 34 25 &   36.6  &    8.9  &   74.8  &    5.4  &   -3    &   22    &   -0.05 &    0.30 &    2.56 &    0.24 &    2.14 &    0.93 & b & GPh\\
W910817  &  17 09 25 &   17.6  &    9.7  &   20.6  &    2.7  &   -7    &   13    &   -0.38 &    0.68 &    4.16 &    0.61 &    8.4  &    2.8  & b & PhU \\
W910821  &  10 33 42 &   73    &   12    &  106.9  &    7.8  &  179    &   34    &    1.68 &    0.34 &    2.53 &    0.26 &    6.1  &    1.1  &   & GPhS\\
W910901  &  05 24 34 &   13    &   11    &   12.6  &    2.7  &   22    &   14    &    1.7  &    1.1  &    0.59 &    0.11 &    1.19 &    0.64 & b & Ph\\
W910925  &  03 35 10 &   45    &   10    &   17.2  &    3.6  &  -18    &   17    &   -1.0  &    1.0  &    0.57 &    0.18 &    0.77 &    0.85 &   & B\\
W910927  &  23 27 00 &   28.7  &    6.2  &   50.6  &    3.5  &   76    &   18    &    1.51 &    0.38 &    4.19 &    0.25 &    8.4  &    1.1  &   & BDU\\
W911016  &  11 01 34 &  264    &   19    &   65.9  &    5.4  &  135    &   45    &    2.06 &    0.71 &    1.07 &    0.12 &    3.46 &    0.92 &   & BDGPPhU\\
W911022  &  04 14 00 &   \multicolumn{2}{c}{--} &    8.9  &    2.7  &    3    &   14    &    0.4  &    1.6  &    0.85 &    0.19 &    2.1  &    1.0 &   & BU\\
W911202  &  20 28 51 &   16.5  &    1.1  &   69.6  &    2.1  &  152    &   13    &    2.19 &    0.21 &    6.24 &    0.52 &   14.8  &    2.9  &   & BDPU\\
W911205  &  23 04 55 &   29.5  &    6.4  &   14.4  &    2.6  &   50    &   20    &    3.5  &    1.5  &    0.70 &    0.15 &    2.3  &    1.1  &   & B\\
W911209  &  18 36 11 &   22.1  &    2.5  &   26.3  &    1.7  &   71.7  &    4.7  &    2.73 &    0.26 &    4.26 &    0.42 &   12.7  &    1.2  &   & BUY\\
W920116  &  18 30 47 &   \multicolumn{2}{c}{--} &    3.0  &    1.3  &    7.9  &    3.4  &    2.6  &    1.5  &    0.41 &    0.11 &    0.53 &    0.30 &   & BP\\
W920130  &  00 28 36 &  287    &   10    &  136.8  &    6.3  &   80    &   15    &    0.59 &    0.11 &    1.31 &    0.15 &    1.73 &    0.35 & c & B\\
W920210  &  09 53 52 &   36.1  &    6.2  &   25.8  &    2.6  &   89.6  &    7.3  &    3.47 &    0.45 &    1.06 &    0.14 &    4.12 &    0.41 &   & BPhS\\
W920302  &  06 17 16 &   58.8  &    5.6  &   12.2  &    2.3  &   21.0  &    6.0  &    1.72 &    0.59 &    0.47 &    0.11 &    0.69 &    0.28 &   & B\\
W920307  &  00 18 11 &   58.8  &    8.0  &   16.9  &    2.0  &   26.5  &    5.4  &    1.57 &    0.37 &    1.29 &    0.11 &    2.21 &    0.27 &   & BU\\
W920311  &  02 20 26 &   29.0  &    6.2  &   86.7  &    4.1  &  252    &   13    &    2.91 &    0.21 &    5.05 &    0.23 &   14.48 &    0.73 &   & BDPPhSU\\
W920325  &  17 17 37 &   51.7  &    3.3  &   31.3  &    1.7  &   50.5  &    4.4  &    1.61 &    0.17 &    2.46 &    0.30 &    7.56 &    0.92 &   & BDPPhSU\\
W920404  &  13 11 45 &    2.5  &    1.0  &    9.19 &    0.95 &   28.1  &    2.6  &    3.06 &    0.42 &    3.67 &    0.42 &   14.3  &    1.3 &   & BU\\
W920419  &  20 07 01 &   26    &   11    &    4.1  &    1.5  &   21.9  &    4.0  &    5.2  &    2.1  &    0.27 &    0.07 &    0.72 &    0.19 &   & BU\\
W920615  &  14 24 58 &   \multicolumn{2}{c}{--}  &    6.7  &    1.7  &    5.9  &    9.5  &    0.8  &    1.4  &    0.79 &    0.17 &    0.76 &    0.93  &   & B\\
W920711  &  16 09 17 &   80.2  &    8.9  &   83.7  &    5.9  &  357    &   34    &    4.27 &    0.52 &    2.11 &    0.20 &   10.0  &    1.2  & b & BDPUY\\
W920714  &  13 04 33 &   28.9  &    5.1  &   30.8  &    1.7  &   81.8  &    3.2  &    2.65 &    0.18 &    1.59 &    0.11 &    4.60 &    0.24 &   & BPPhSU\\
\end{tabular}
\end{table*}

\setcounter{table}{0}
\begin{table*}
{\bf Table~1}~(continued)
\vspace{3mm}
\begin{tabular}{llr@{$\pm$}lr@{$\pm$}lr@{$\pm$}lr@{$\pm$}lr@{$\pm$}lr@{$\pm$}lll}
\noalign{\smallskip}
\hline
\noalign{\smallskip}
Burst & Start & \multicolumn{2}{c}{$T_{90}$} & \multicolumn{4}{c}
{\mbox{Fluence (10$^{-7}$ erg/cm$^{2}$})} & \multicolumn{2}{c} {Hardness} &
\multicolumn{4}{c} {\mbox{Peak flux (10$^{-7}$ erg/cm$^{2}$/s})} & Rem & Confirm$^{\rm a}$\\ 
\cline{5-8} \cline{11-14}  
name & (UT) & \multicolumn{2}{c} {(s)} & \multicolumn{2}{c} {8--20 keV} &
\multicolumn{2}{c} {20--60 keV} & \multicolumn{2}{c} {ratio} &
\multicolumn{2}{c} {8--20 keV} & \multicolumn{2}{c} {20--60 keV} & \\    
\hline
\noalign{\smallskip}
W920718  &  14 40 43 &   64.6  &    8.8  &   58.4  &    2.5  &   26    &   10    &    0.46 &    0.18 &    2.02 &    0.11 &    2.40 &    0.39 &   & BU\\
W920718b &  21 32 44 &    4.50 &    0.80 &   12.92 &    0.74 &   29.4  &    1.5  &    2.28 &    0.18 &    3.69 &    0.32 &    9.21 &    0.73 &   & BPhU\\
W920720  &  05 53 20 &  117    &   20    &   41.9  &    3.2  &   94    &   10    &    2.25 &    0.29 &    0.72 &    0.10 &    1.69 &    0.30 &   & BPU\\
W920723  &  01 00 49 &   32.7  &    8.0  &   \multicolumn{2}{c}{--}  &
\multicolumn{2}{c}{--} &  \multicolumn{2}{c}{--} &  \multicolumn{2}{c}{--} &  \multicolumn{2}{c}{--}  & d & BDPhU\\
W920723b &  20 03 09 &   53.0  &    3.7  &   59.4  &    2.1  &  209.9  &    4.3  &    3.54 &    0.15 &    7.83 &    0.41 &   41.1  &    1.2  &   & PPhSU\\
W920814  &  06 10 35 &   21.6  &    6.2  &   10.5  &    1.3  &   19.9  &    2.5  &    1.89 &    0.34 &    0.56 &    0.08 &    1.09 &    0.16 &   & BEU\\
W920902  &  00 29 02 &    7.3  &    1.1  &   11.7  &    1.1  &   85.2  &    3.0  &    7.26 &    0.76 &    2.02 &    0.35 &   15.7  &    1.1  &   & BDPhU\\
W920903  &  01 35 46 &  131.8  &    6.3  &   42.0  &    3.4  &  107.4  &    7.3  &    2.56 &    0.27 &    0.58 &    0.10 &    1.80 &    0.22 &   & PhU\\
W920903b &  23 29 01 &   23.8  &    1.9  &   61.5  &    2.4  &  171.3  &    6.0  &    2.78 &    0.15 &    6.99 &    0.45 &   27.7  &    1.3  &   & PhU\\
W920925  &  20 30 42 &   57.1  &    6.8  &   22.2  &    2.7  &   44.5  &    5.5  &    2.00 &    0.35 &    2.23 &    0.38 &    8.13 &    0.97 &   & PhU\\
W920925b &  21 45 20 &   12.9  &    5.6  &   10.8  &    1.3  &   20.2  &    3.8  &    1.87 &    0.42 &    1.00 &    0.12 &    2.65 &    0.36 &   & BU\\
W920925c &  22 46 24 &  282.1  &    8.8  &  120.3  &    4.8  &  207    &   10    &    1.73 &    0.11 &    1.55 &    0.11 &    3.50 &    0.24 &   & U\\
W921001  &  04 03 54 &  \multicolumn{2}{c}{--}  &    4.8  &    1.3  &   13.6  &    2.7  &    2.80 &    0.96 &    0.73 &    0.30 &    1.80 &    0.63 &   & B\\
W921013  &  20 32 32 &   12.9  &    2.0  &    6.50 &    0.85 &   15.9  &    1.9  &    2.46 &    0.44 &    1.15 &    0.21 &    5.92 &    0.61 &   & BU\\
W921013b &  23 00 42 &  344.6  &    3.7  &  114.2  &    4.9  &  209    &   11    &    1.84 &    0.13 &    8.11 &    0.43 &   26.8  &    1.2  &   &  PhU\\
W921021  &  18 20 46 &   32.4  &    7.9  &   14.6  &    2.0  &   24.8  &    5.9  &    1.70 &    0.47 &    0.82 &    0.13 &    1.56 &    0.36 &   & B\\
W921022  &  15 21 00 &   95    &   13    &   57.2  &    2.8  &   49.8  &    5.9  &    0.87 &    0.11 &    1.84 &    0.30 &    4.95 &    0.72 &   & BEUY\\
W921025  &  13 55 40 &   21.6  &    8.8  &    4.3  &    1.9  &   20.3  &    3.8  &    4.7  &    2.2  &    0.36 &    0.10 &    1.04 &    0.21 & b & PhU\\
W921029  &  12 38 05 &   80.1  &    8.9  &   18.8  &    2.1  &   27.1  &    5.0  &    1.44 &    0.31 &    0.90 &    0.09 &    1.18 &    0.18 &   & BU\\
W930106  &  15 37 40 &    0.06 &    0.03 &   \multicolumn{2}{c}{--}  &    \multicolumn{2}{c}{--} &  \multicolumn{2}{c}{--} &  \multicolumn{2}{c}{--}  &  \multicolumn{2}{c}{--}  & d & BDPhSUY\\
W930116  &  02 47 06 &    7.3  &    5.1  &    3.9  &    1.2  &   13.0  &    2.9  &    3.3  &    1.3  &    0.37 &    0.12 &    1.24 &    0.28 &   & BU\\
W930609  &  10 07 30 &   50    &   15    &    6.4  &    1.9  &   18.2  &    3.6  &    2.8  &    1.0  &    0.28 &    0.09 &    0.96 &    0.18 &   & BU\\
W930612  &  00 44 20 &  130    &   51    &   29.9  &    3.1  &   77.0  &    7.5  &    2.57 &    0.37 &    1.00 &    0.09 &    3.48 &    0.23 &   & BCDMPhU\\
W930703  &  11 26 30 &    7.3  &    1.1  &   13.5  &    1.6  &   37.3  &    3.2  &    2.75 &    0.41 &    3.01 &    0.52 &   16.4  &    1.3  &   & \\
W930705  &  12 39 18 &  152    &   29    &   36.1  &    3.8  &   25.3  &    7.4  &    0.70 &    0.22 &    0.45 &    0.08 &    0.99 &    0.16 &   & BU\\
W930706  &  05 13 31 &    4.50 &    0.80 &   16.71 &    0.81 &   37.8  &    1.6  &    2.26 &    0.15 &    8.48 &    0.42 &   24.8  &    1.0  &   & BMPhU\\
W930710  &  03 14 42 &   14.4  &    6.2  &    4.5  &    1.3  &   12.8  &    3.0  &    2.8  &    1.0  &    0.42 &    0.11 &    0.87 &    0.24 &   & BU\\
W930714  &  16 13 04 &   14.4  &    8.8  &   \multicolumn{2}{c}{--}  &    \multicolumn{2}{c}{--} &  \multicolumn{2}{c}{--} &  \multicolumn{2}{c}{--}  &  \multicolumn{2}{c}{--}  & d & BPhU\\
W930910  &  12 12 30 &   79.8  &    6.2  &   27.8  &    6.5  &   78    &   14    &    2.82 &    0.85 &    0.88 &    0.26 &    2.74 &    0.59 &   & BU\\
W930927  &  04 18 15 &   50    &   11    &   11.4  &    1.6  &   25.4  &    4.7  &    2.21 &    0.52 &    0.59 &    0.08 &    1.70 &    0.23 &   & BU\\
W940307  &  07 56 29 &   14.3  &    8.7  &    2.9  &    1.4  &    8.5  &    2.6  &    2.8  &    1.6  &    0.18 &    0.08 &    0.59 &    0.16 &   & B\\
W940329  &  18 15 44 &   72    &   11    &   \multicolumn{2}{c}{--}  &
   \multicolumn{2}{c}{--} &  \multicolumn{2}{c}{--} &  \multicolumn{2}{c}{--}  &  \multicolumn{2}{c}{---}   & d & BU\\
W940419  &  19 11 07 &   84    &   10    &   59.5  &    3.5  &  213    &   12    &    3.59 &    0.30 &    2.02 &    0.14 &    6.21 &    0.49  &   & BPhU\\
W940619  &  21 32 32 &   72    &   11    &   \multicolumn{2}{c}{--}  &    \multicolumn{2}{c}{--} &  \multicolumn{2}{c}{--} &  \multicolumn{2}{c}{--}  &  \multicolumn{2}{c}{--}   &   & BCPhU\\
W940701  &  21 44 29 &  201    &   23    &   54.9  &    4.2  &  307    &   16    &    5.61 &    0.53 &    1.18 &    0.11 &    2.51 &    0.39 & c & BU\\
W940703  &  04 40 55 &   45.6  &    7.9  &  239.2  &    4.6  & 1010    &   17    &    4.22 &    0.11 &    9.36 &    0.22 &   40.46 &    0.93 &   & BPhSUY\\
W940711  &  18 14 10 &   32.4  &    5.6  &   11.1  &    1.9  &    0.2  &    6.6  &    0.03 &    0.60 &    0.53 &    0.11 &    0.53 &    0.40 &   & B\\
W940907  &  20 36 20 &   26    &   11    &   20.9  &    2.6  &  115    &   10    &    5.50 &    0.85 &    0.95 &    0.10 &    5.62 &    0.43 &   & \\
W940910  &  23 57 56 &   51    &   15    &   38.1  &    3.8  &   51    &   13    &    1.36 &    0.37 &    1.09 &    0.11 &    2.30 &    0.40 &   & Ph\\
\noalign{\smallskip}
\hline
\end{tabular}

\smallskip
{\it Notes:}\\
$^{\rm a}$ Bursts were also observed by other instruments: B-BATSE/CGRO,
C-COMPTEL/CGRO, D-DMS, E-WATCH/EURECA, G-GINGA, K-KONUS/GRANAT, M-Mars
Observer, O-OSSE/CGRO, P-PVO, Ph-PHEBUS/GRANAT, S-SIGMA/GRANAT, U-ULYSSES,
Y-YOHKOH.

$^{\rm b}$ Direction of burst arrival unknown. 

$^{\rm c}$ Burst superimposed on variable background. Values given for
the parameters possibly in error.  

$^{\rm d}$ Burst source was either outside or at the border of the WATCH field
of view according to the BATSE localization (Meegan et al., 1996).
\end{table*}

\setcounter{figure}{1}
\begin{figure*}[p]
\epsfxsize=19cm \epsfbox{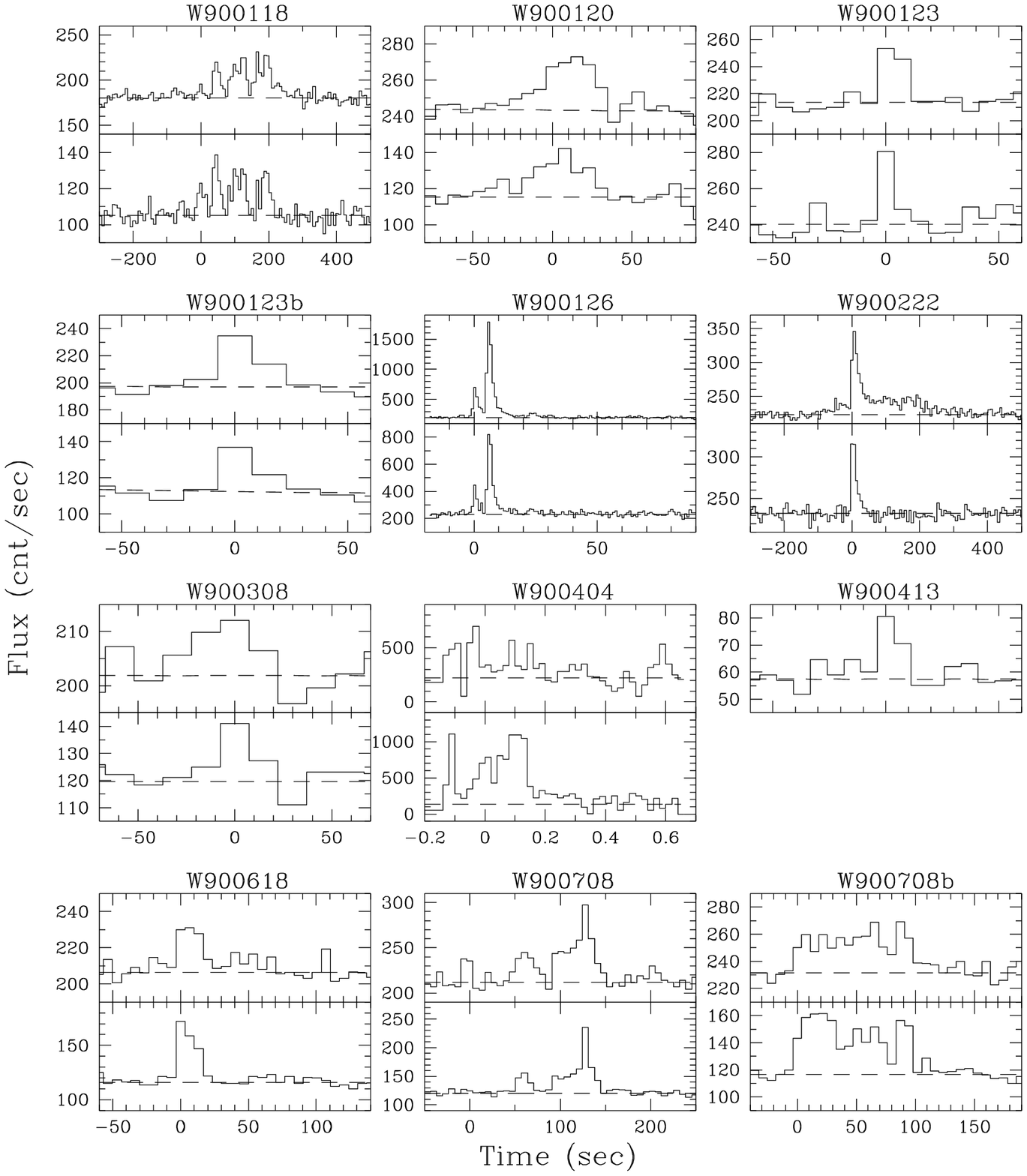}
\caption[]{Time histories of the gamma-ray bursts detected by WATCH
in the two energy ranges: 8--20~keV (upper panels) and  20--60~keV
(lower panels). Time runs relative to the burst start (see
Table~1). The background count rate is indicated by the dashed line}
\label{fig2}
\end{figure*}

\setcounter{figure}{1}
\begin{figure*}
\epsfxsize=19cm \epsfbox{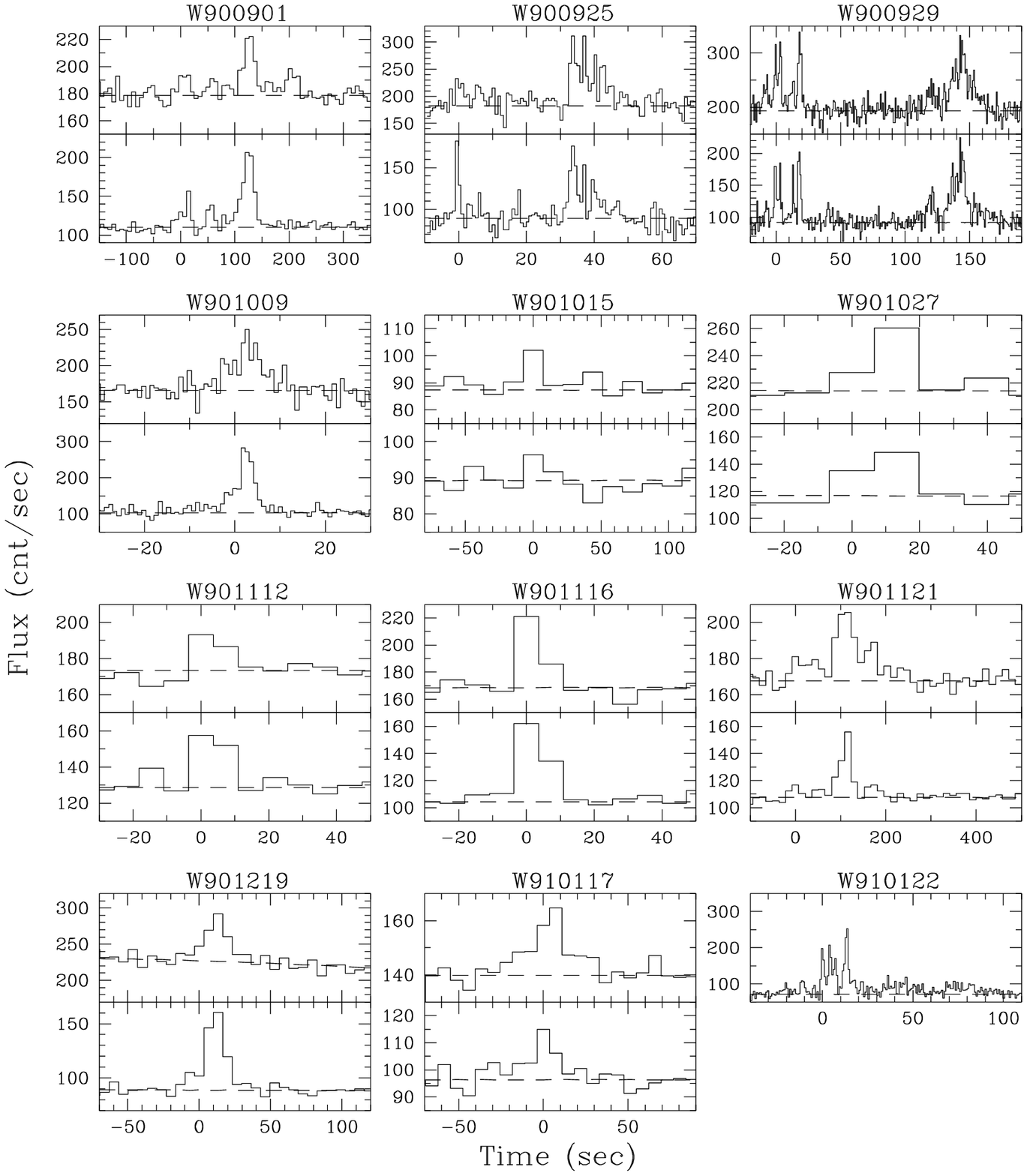}
\caption[]{continued}
\end{figure*}

\setcounter{figure}{1}
\begin{figure*}
\epsfxsize=19cm \epsfbox{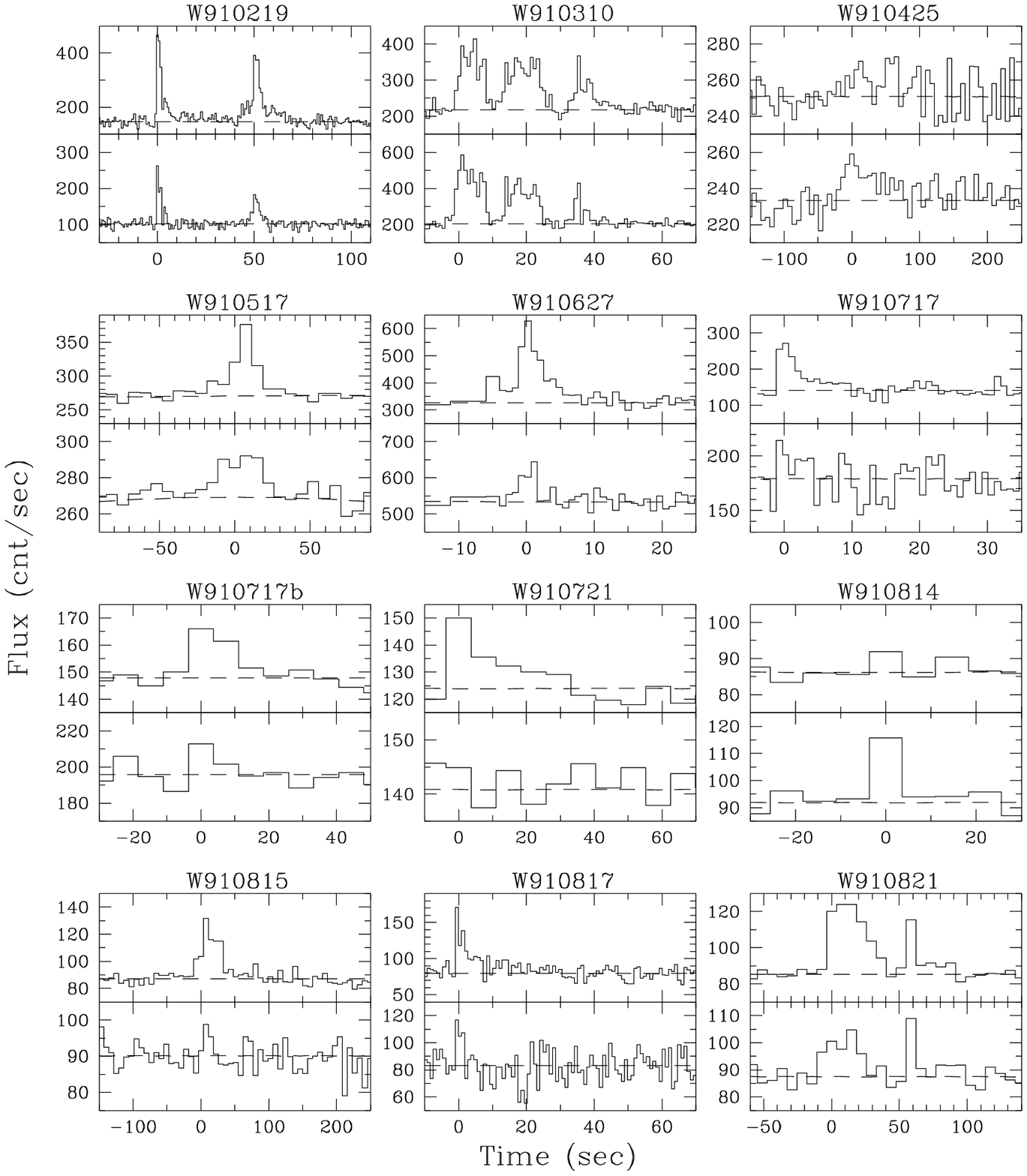}
\caption[]{continued}
\end{figure*}

\setcounter{figure}{1}
\begin{figure*}
\epsfxsize=19cm \epsfbox{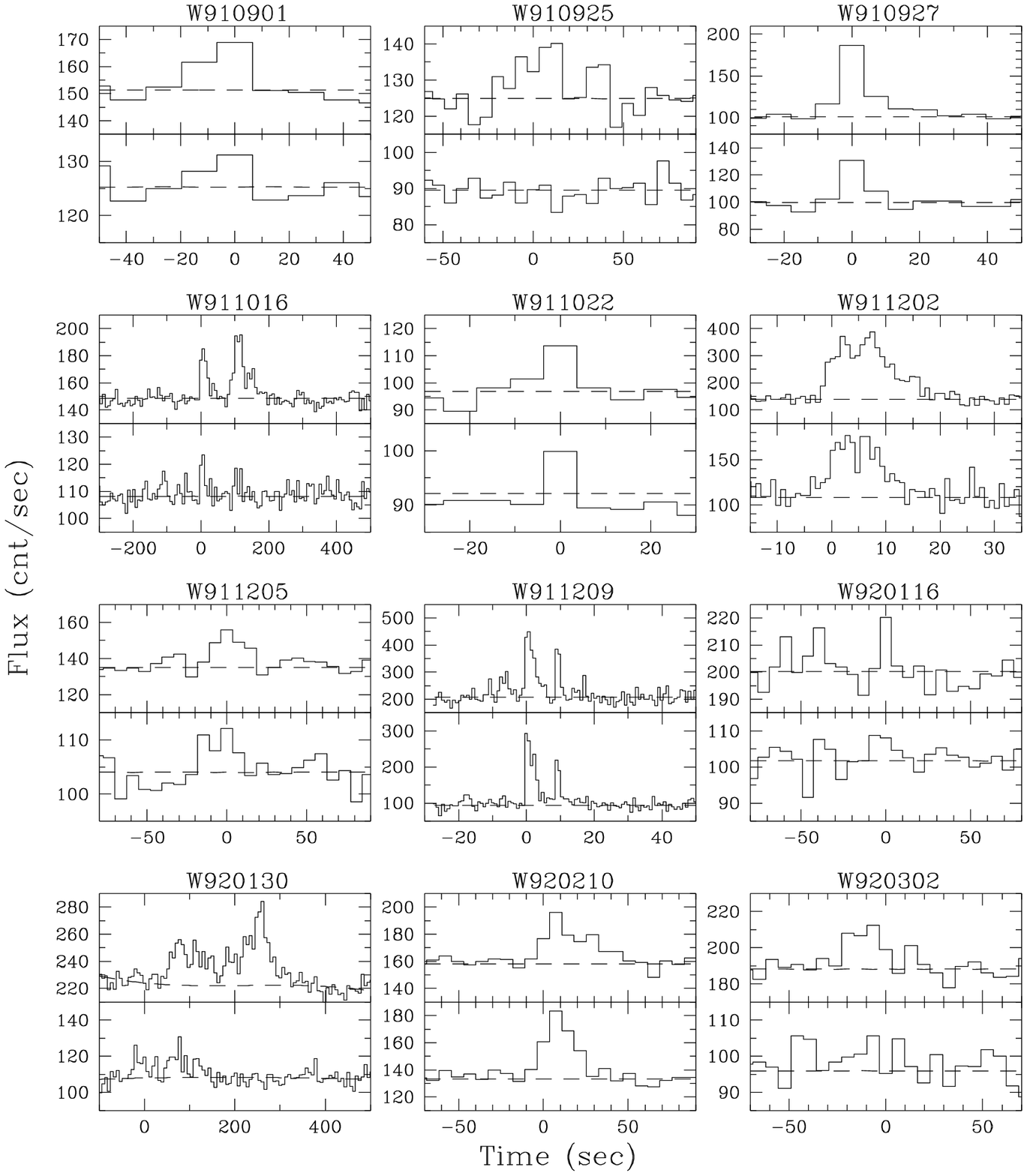}
\caption[]{continued}
\end{figure*}

\setcounter{figure}{1}
\begin{figure*}
\epsfxsize=19cm \epsfbox{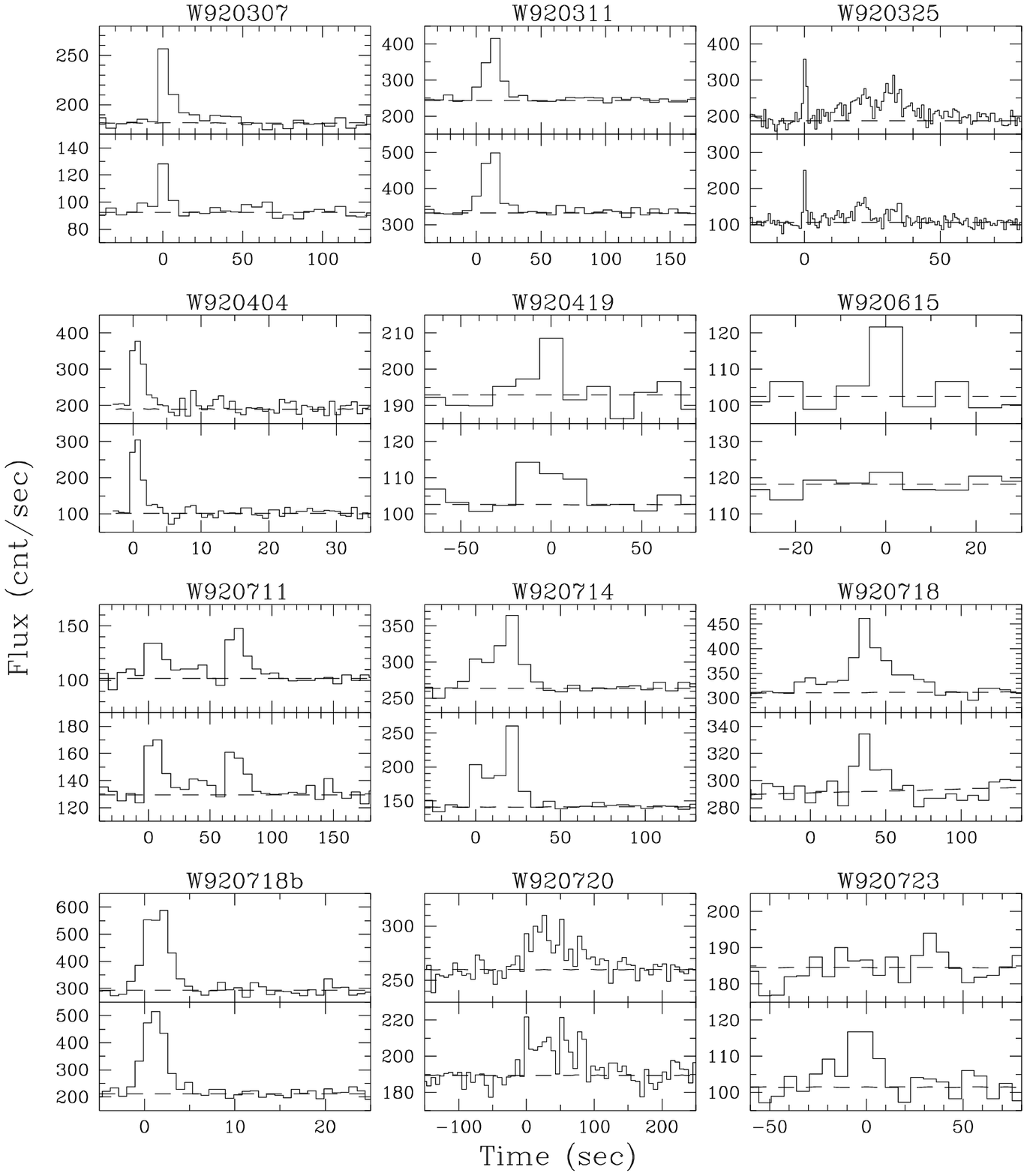}
\caption[]{continued}
\end{figure*}

\setcounter{figure}{1}
\begin{figure*}
\epsfxsize=19cm \epsfbox{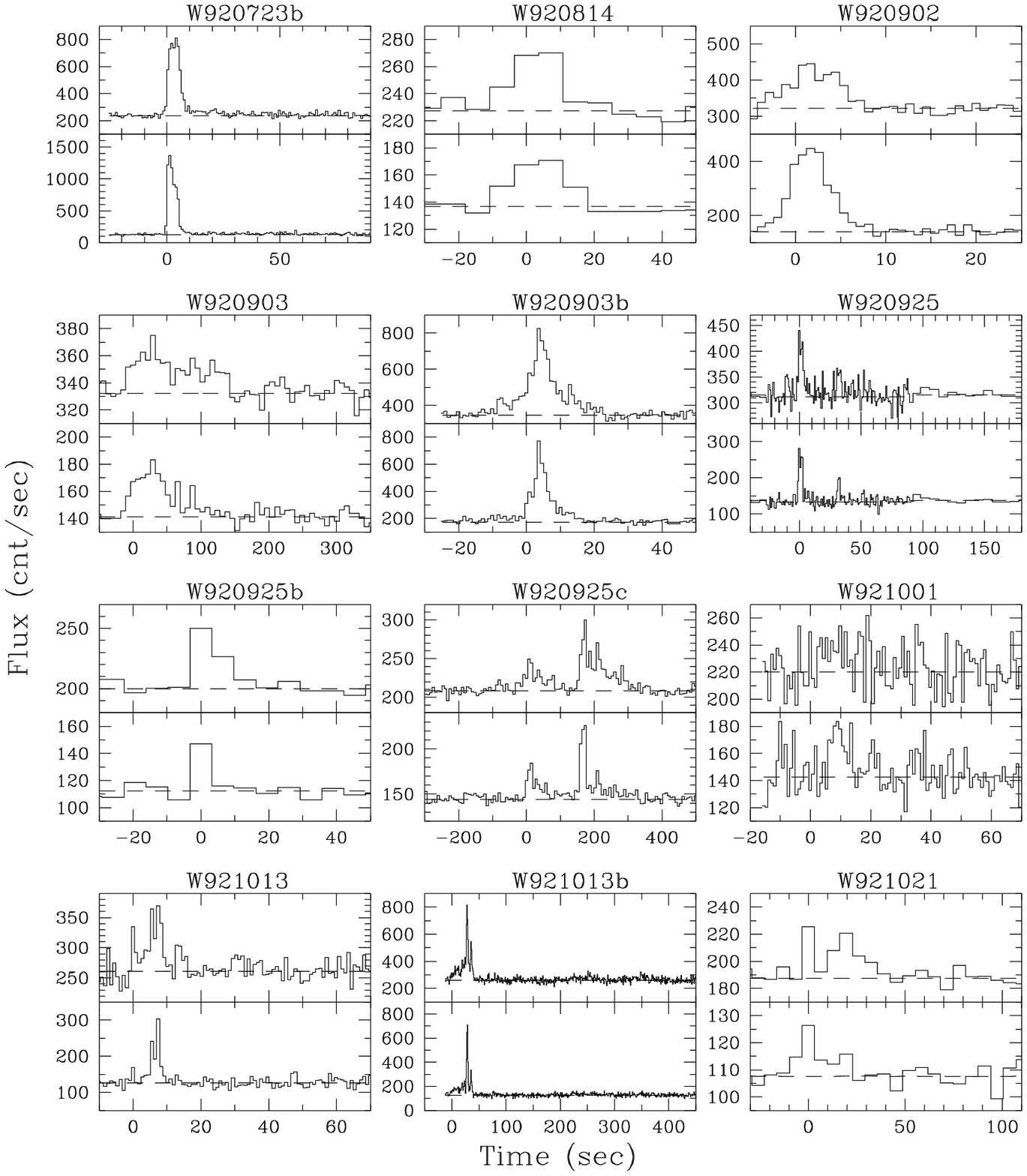}
\caption[]{continued}
\end{figure*}

\setcounter{figure}{1}
\begin{figure*}
\epsfxsize=19cm \epsfbox{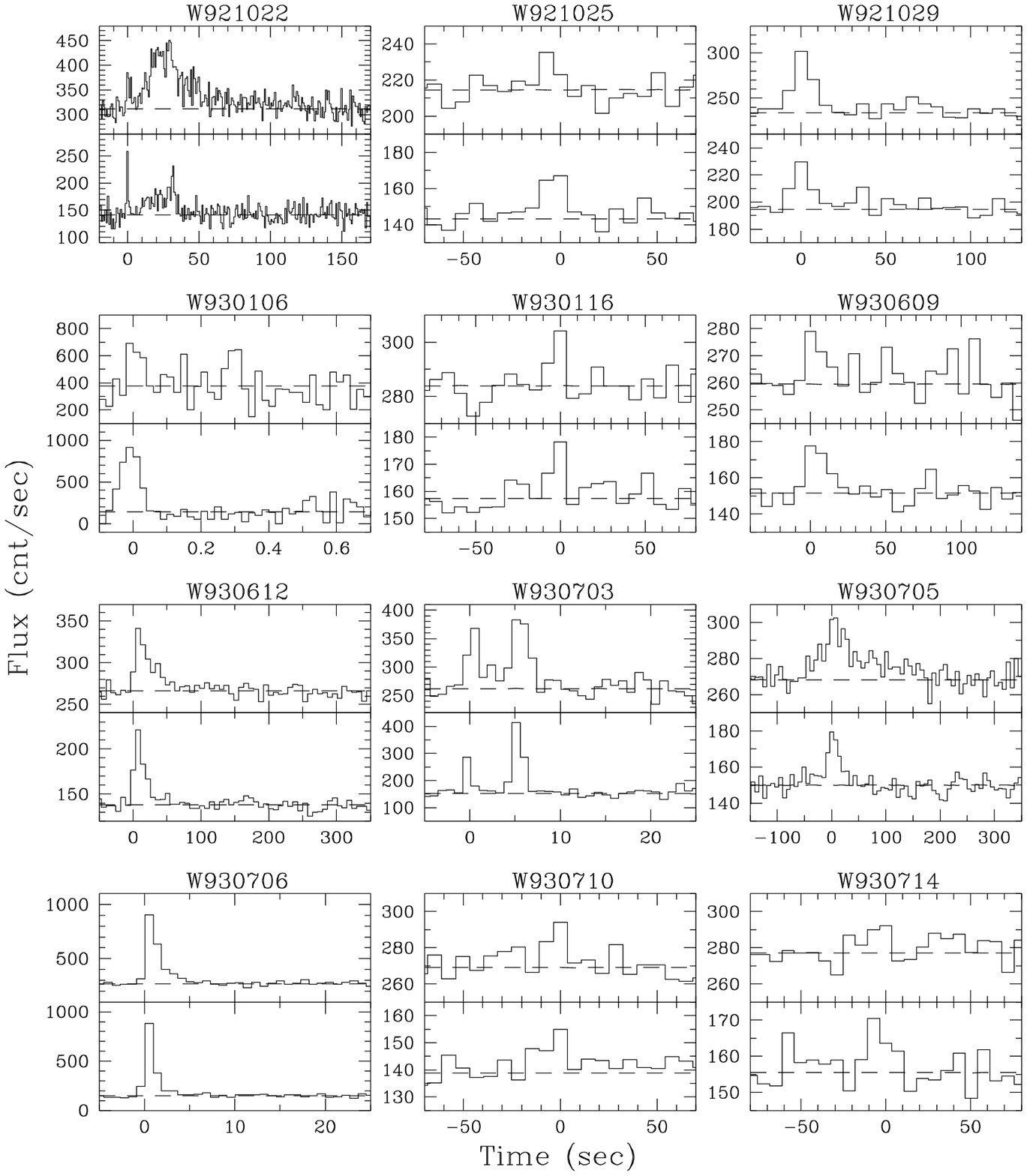}
\caption[]{continued}
\end{figure*}

\setcounter{figure}{1}
\begin{figure*}
\epsfxsize=19cm \epsfbox{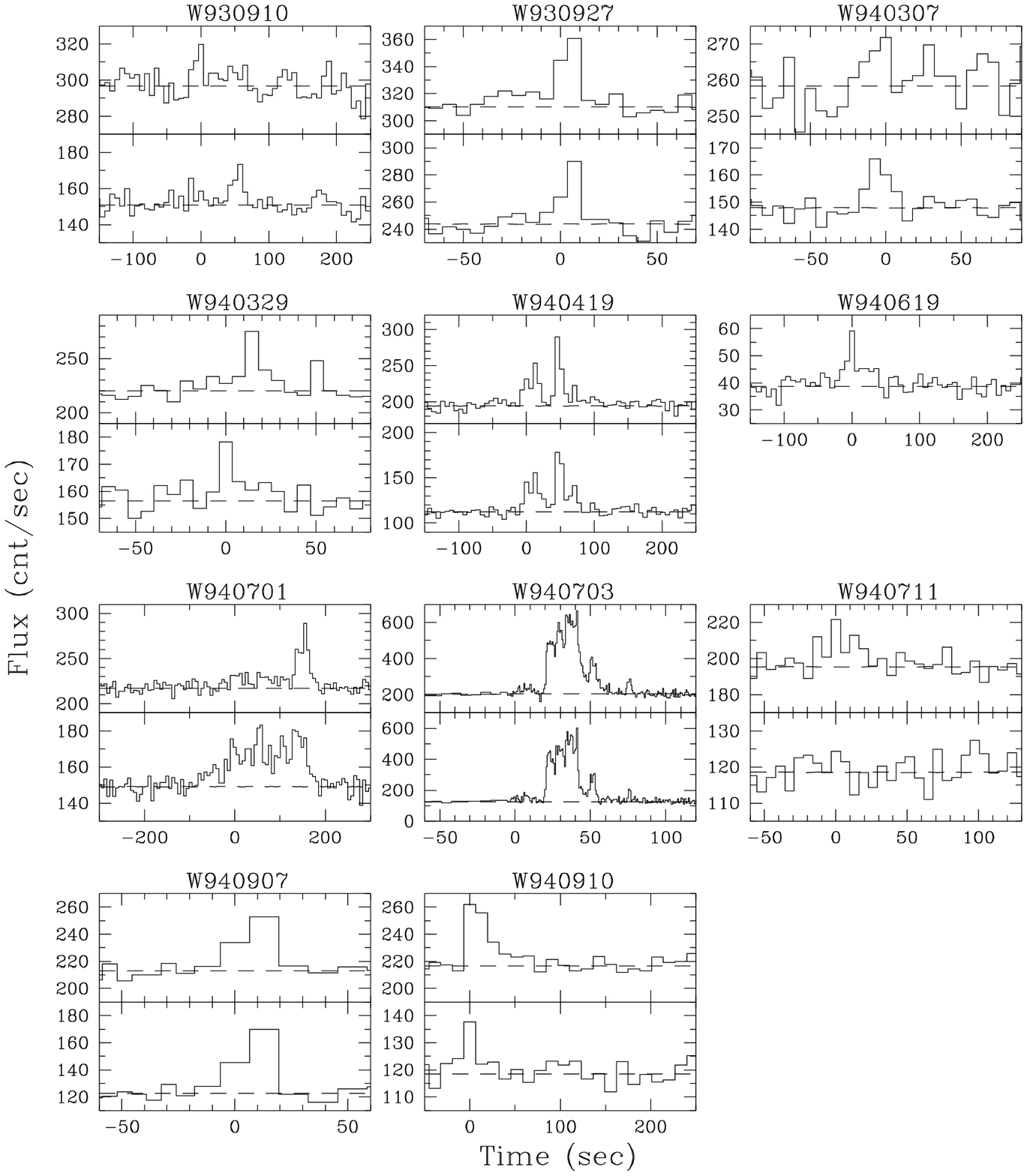}
\caption[]{continued}
\end{figure*}

\end{document}